# Gate-Defined Quantum Dots on Carbon Nanotubes

M. J. Biercuk, S. Garaj, N. Mason, J. M. Chow, C. M. Marcus
*Department of Physics, Harvard University*
*Cambridge, MA 02138*

**We report the realization of nanotube-based multiple quantum dots that are fully defined and controlled with electrostatic gates. Metallic top-gates are used to produce localized depletion regions in the underlying tubes; a pair of such depletion regions in a nanotube with ohmic contact electrodes defines the quantum dot. Top-gate voltages tune the transparencies of tunnel barriers as well as the electrostatic energies within single and multiple dots. This approach allows precise control over multiple devices on a single tube, and serves as a design paradigm for nanotube-based electronics and quantum systems.**

A number of proposed solid-state devices[1] take as their fundamental element the quantum dot—a classically isolated island of electrons with a discrete energy spectrum [1, 2]. As a substrate for realizing multiple quantum-dot devices, carbon nanotubes [3] offer a variety of appealing physical properties. However, nanotube-based electronics in general have been limited by the difficulty of fabricating complex devices on a single tube. In previous studies, isolated quantum dots formed in carbon nanotubes were defined either by tunnel barriers at the metal-nanotube interface [4, 5], or by intrinsic [6, 7] or induced [8, 9] defects along the tube. These devices demonstrated the potential of nanotube-based quantum devices but did not allow independent control over device parameters (e.g., charge number and tunnel barrier transparency), and also placed stringent geometric constraints on device design. In the present study, we address some of these challenges by forming the quantum dots on the nanotube using only patterned gates, while the contacts to the nanotube remain highly transparent. This design allows multiple quantum dots to be arbitrarily positioned along a tube (quantum dots connected to 1D nanotube leads), with independent control over tunnel barriers and dot charges. A backgate is used to set overall carrier density. Here, we show that quantum dots fabricated in this manner exhibit familiar characteristics, yet provide significant advances in device control. In particular, full control over tunnel barrier locations and transparencies should allow improvements in the study and control of spin and charge dynamics in carbon nanotubes.

Nanotubes were grown via chemical vapor deposition from lithographically defined Fe catalyst islands on a degenerately doped Si wafer with 1μm of thermally grown oxide (See Fig. 1a). Atomic force microscopy was used to locate nanotubes relative to alignment markers, and single-walled tubes with diameters less than ~ 3 nm were contacted with 15 nm of Pd, patterned by electron beam lithography [10]. Device lengths were in the range 5–25 μm. After contacting, the entire sample was coated with 25–35 nm of either $SiO_2$ deposited by plasma-enhanced chemical vapor deposition (PECVD) or $Al_2O_3$ deposited by atomic layer deposition (ALD). Cr/Au top-gates, 150–300 nm wide, were then patterned over the tubes using electron-beam lithography, with care taken to prevent overlap between the gates and the Pd contacts.

Two-terminal conductance was measured in either a pumped $^4$He cryostat (300K to 1.5K) or a $^3$He cryostat (300K to 0.3K). Current and differential conductance, $dI/dV$, were measured simultaneously using a combined ac + dc voltage bias with a current amplifier (Ithaco 1211) and lockin amplifier.

All semiconducting nanotubes coated with PECVD SiO$_2$ showed p-type field effect transistor behavior at room-temperature, exhibiting carrier depletion with positive voltage applied to top or back-gates (conductance was typically suppressed by three to four orders of magnitude at voltages of ~ 1 – 3 V) [11].  In contrast, some tubes coated with ALD showed ambipolar behavior (high conductance at both positive and negative applied top or back-gate voltage surrounding a low-conductance region) with thermally activated conductance in the gap.  We believe this difference may be due to oxygen doping of the nanotubes by the highly energetic oxygen plasma in the PECVD process.  Room-temperature maximum conductances at zero dc bias, $V_{SD} = 0$, ranged from ~ 0.5 – 2 $e^2/h$.  Overall, devices coated with PECVD and ALD were qualitatively similar in behavior.  Devices showing a relatively weak gate response (presumably metallic) were not investigated further.

Single quantum dots were formed using a three-gate configuration (Fig. 1a), where outer gates act as tunnel barriers defining the dot (denoted "barrier 1" and "barrier 2") by locally depleting carriers beneath them, and a center gate (denoted "plunger") shifts the chemical potential in the dot relative to the chemical potentials of the contacts and the segments of the tube away from the gates. Gate response of tube conductance (Fig. 1b) for a single-dot device using PECVD SiO$_2$ (micrograph of similar device shown in Fig. 1b, inset; the two center gates are connected and act as one plunger gate) shows p-type field effect behavior, that is, the tube is depleted when any top-gate bias becomes sufficiently positive. The gated region of the nanotube is ~ 2 μm in length, while the total tube length is ~ 25 μm between the Pd contacts.

The independent action of the barrier gates is evident in the two-dimensional plot of differential conductance, $dI/dV$, as a function of barrier gate voltages, shown in Fig. 2a. The square edge of the conducting region demonstrates independent barrier depletion with little cross-coupling. Parallel diagonal features separated by ~2 mV visible near the pinch-off of both barrier gates (Fig. 2b) are a signature of Coulomb blockade, discussed below. Figure 2b demonstrates that the barrier gate voltages can be controllably adjusted to allow a transition from open conduction to weak tunneling through a gate-defined quantum dot resulting in the appearance of Coulomb blockade peaks.

Examined as a function source-drain voltage bias, $V_{SD}$, as well as plunger gate voltage, the Coulomb blockade peaks form a series of repeated "Coulomb diamonds" where conduction is suppressed whenever the energy to add the next hole to the device exceeds $V_{SD}$ (Fig. 2c). The ratio, $\eta^{-1} \sim 0.85$, of Coulomb diamond height ($V_{SD}$) to width (plunger voltage) gives the conversion, $e\eta$, from the distance in plunger voltage between Coulomb blockade peaks to the dot charging energy $E_C = e^2/C \sim e^2/\kappa\varepsilon_0 L$ ($\kappa \sim 4$ is the dielectric constant of $SiO_2$), and indicates a strong coupling of the plunger gate to the dot [12]. From this analysis, one extracts a dot length of $L \sim 2$ μm, comparable to the length of the gated region of the tube, and much less than the ~ 25 μm tube length. Several hundred consecutive Coulomb peaks are visible over a range of plunger gate voltage > 4V. Throughout this range, peak heights remain controllable by adjusting the barrier gates.

The data in Fig. 2c show significant off-resonance tunneling outside of the Coulomb diamonds, obscuring any excited state features that would be expected in low-temperature transport. We find that this washing out of excited state features is characteristic of devices made with PECVD $SiO_2$ but is typically not the case for ALD $Al_2O_3$, where excited state features are generally visible outside the Coulomb diamonds. We do not know if this difference is due to the oxide material itself or due to damage that occurs during deposition. Figure 3 shows a series of Coulomb diamonds measured on a gate-defined quantum dot using ALD $Al_2O_3$ as an insulating layer. Off-resonance conductance is low, and excited states are visible outside of the boundaries of the diamonds. The mean level spacing extracted from the data $\Delta E = hv_F / 2L \sim$ 2-3 mV gives a measure of dot length, $L \sim$ 0.5-0.8 μm, again roughly consistent with the lithographic dimensions of the gated region of the tube for this device.

Taking advantage of the versatility of gate-defined devices, we next investigate a double quantum dot formed by three depletion regions along a nanotube. The double quantum dot, shown in Fig. 4a, comprises left and right barrier gates, a middle barrier gate and two independent plunger gates. Gates are approximately 150 nm wide with 150 nm spacing; total nanotube length is ~10 μm, much longer than the double dot.

Resonant transport through double dots in series occurs only when available energy levels in each dot align with each other and with the chemical potentials in the two leads [13]. When the mutual capacitance of the dots is weak, the alignment condition occurs at the intersection of the Coulomb peaks leading to a rectangular grid of resonant conduction peaks. Such a pattern is seen in Fig. 4b. The resulting charge stability diagram forms approximately square cells, each corresponding to a fixed charge number in both dots. In the regime shown in Fig. 4b, cross-coupling of plunger gates, which would skew the square pattern into rhombus shapes, appears to be quite small. We note that these measurements were conducted at $T \sim 1.5$K, higher than the typical temperature where double dots based on semiconductor heterostructures have been measured [13].

Gate voltages can also be used to control interdot coupling, allowing a transition from two isolated dots (uncoupled) to one large dot (fully coupled). Increasing the coupling between the two dots (by reducing the voltage on the middle gate) leads to a splitting of the high-conductance points of degeneracy between different charge configurations, and the emergence of a honeycomb pattern in the charge stability diagram (Fig. 4c). The interdot interaction may be due to capacitive or tunnel-coupling, but it has been shown that tunnel-coupling increases exponentially faster with a reduction of the interdot barrier [14]. In addition, as the voltage on the middle gate is reduced we observe an increase in vertex height by approximately an order of magnitude. It is therefore likely that finite tunnel coupling leads to the splitting of the vertices [15, 16] in Fig. 4c, which in this case is partially obscured by thermal broadening. (Any splitting of the charge degeneracy points in Fig. 4b is smaller than the thermal smearing.) In addition to the vertex splitting in Fig. 4c, we observe strong conductance along the edges of the honeycomb cells, due to higher-order processes through virtual states. Further decreasing the voltage on the middle gate yields a series of straight diagonal lines as a function of the two plunger gates, as expected when the two dots merge to form a single large dot. The lines arise from Coulomb charging where both gates act additively in coupling to the single particle states of the dot (Fig. 4d) [17]. An analysis of the spacing between the Coulomb peaks along the total energy axis (a line perpendicular to the sloped lines) shows that the peak spacing in Fig. 4d, 4.8 mV, is half of that required to move from one degeneracy point to the next along a similar line in Fig. 4b, 9.6 mV. This factor of two corresponds nearly exactly to the difference in the size of a single dot between the isolated and strongly coupled cases for this device, as peak spacing is inversely proportional to dot length [4]. The fact that a single dot can be formed by outside barriers, with three nominally inactive gates between them, suggests that there is little inadvertent depletion or tube damage from the deposition of these top gates.

Conductance through the double dot for different values of middle gate voltage is consistent with the interpretation that the interaction between the two dots is due to tunnel coupling. If we assume that tunneling rates through the outer barriers, $\Gamma_{BL}$ and $\Gamma_{BR}$, are equal and remain roughly constant in the three coupling regimes (these rates do change somewhat as indicated by a varying value of peak conductance in Fig. 4d, and limiting the validity of the following analysis) we can relate the tunneling rates to the typical peak current for the single dot case as $I_P \sim 80 \text{ pA} = (4V_E e/k_B T)(\Gamma_{BL}\Gamma_{BR}/\Gamma_{BL}+\Gamma_{BR})$ to find $\Gamma_{BL} = \Gamma_{BR} = \Gamma \sim 9$ GHz, where $V_E$ is the excitation voltage used in the measurement, and $T = 1.5$K the measurement temperature (this formulation holds approximately for the conditions met in this case, $h\Gamma < k_B T$) [18]. Using these values we solve for the middle barrier tunnel rate, $\Gamma_M$, from the peak current in the intermediate ($I_P \sim 60$ pA) and weak ($I_P \sim 10$ pA) coupling cases using $\Gamma_M^2 = I_P \Gamma^2 / 4(2e\Gamma - 3I_P)$ for the case of resonant tunneling [19]. We find that for the intermediate coupling $\Gamma_M \sim 700$ MHz, and for the weak coupling case $\Gamma_M \sim 200$ MHz, consistent with a modest visible splitting of the vertices for intermediate coupling. Finally, if we assume that in the weak coupling case tunneling through the middle barrier provides the dominant component of resistance, setting $\Gamma_M = I_P/e$ gives $\Gamma_M \sim 60$ MHz, smaller than, but comparable to the value of 200 MHz obtained above.

The gate-controlled transition from open conduction to the Coulomb blockade regime has been investigated in eight devices with various gate dimensions, configurations, and dot sizes. Double quantum dot devices have been investigated using both $SiO_2$ and ALD $Al_2O_3$. Again, off-resonant conduction is reduced in double dots fabricated with ALD $Al_2O_3$ as a gate oxide compared to those using $SiO_2$. In addition, structure is visible inside the honeycomb vertex triangles for $Al_2O_3$ devices when measured at finite $V_{SD}$.

We note that while nanotube double quantum dots have been previously investigated [11], and showed behavior comparable to the present results, this work differs from previous work in not relying upon intrinsic defects or tunnel barriers at the metal nanotube interface, instead defining quantum dots only using electrostatic gates. Controlled gating at arbitrary points along the tube greatly enhances the functionality of nanotube devices for potential applications ranging from bucket-brigade devices to quantum coherent logic elements. This approach may be particularly useful for quantum logic, as recent theoretical work has shown that a one-dimensional array of coupled quantum dots can be used for quantum computation [20, 21]. Further, the expected long spin coherence lifetime [22] for electrons and holes in nanotubes makes this an attractive material for developing spin-based quantum information storage and processing systems.

This work was supported by ARO/ARDA (DAAD19-02-1-0039 and -0191), NSF-NIRT (EIA-0210736), and the Harvard Center for Nanoscale Systems. M.J.B. acknowledges support from an NSF Graduate Research Fellowship and an ARO-QCGR Fellowship. S. G. acknowledges support from Bourse de Recherche, Swiss National Science Foundation. N. M. acknowledges support from the Harvard Society of Fellows.

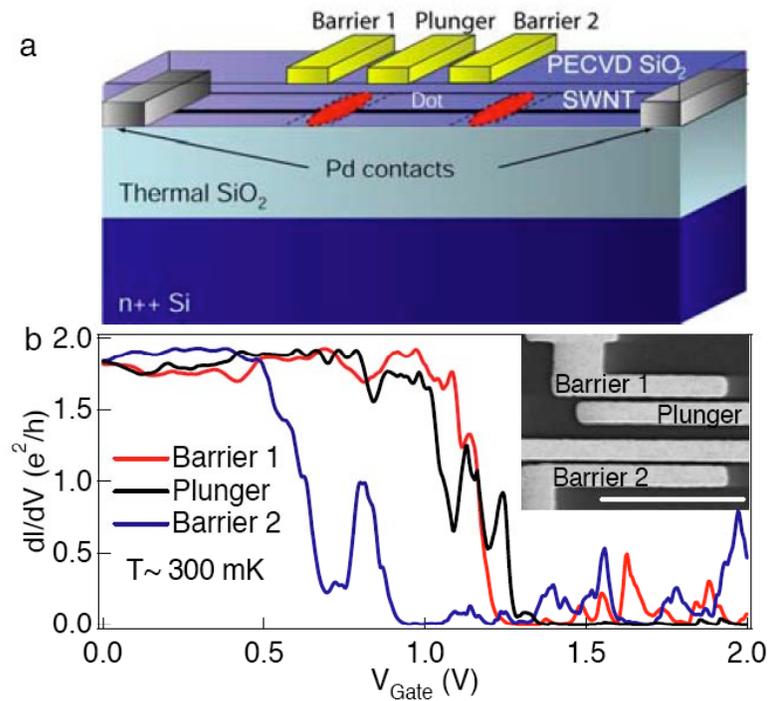

Figure 1a.  Schematic of a gate-defined carbon nanotube quantum dot showing vertically integrated geometry and ohmic contacts. Pd provides high-conductance contacts at the metal-nanotube interface which do not form tunnel barriers at low temperatures.

1b.  Gate response of a ~25 μm long nanotube contacted with Pd, top-gated using PECVD SiO$_2$ at T ~ 300mK with ~10 μV ac excitation.  For this device, all gates strongly suppress conductance at voltages above ~ +1V.  Inset:  SEM of a lithographically similar gate pattern.  The middle two gates are connected together and serve as a single plunger gate.  Scale bar = 2 μm.

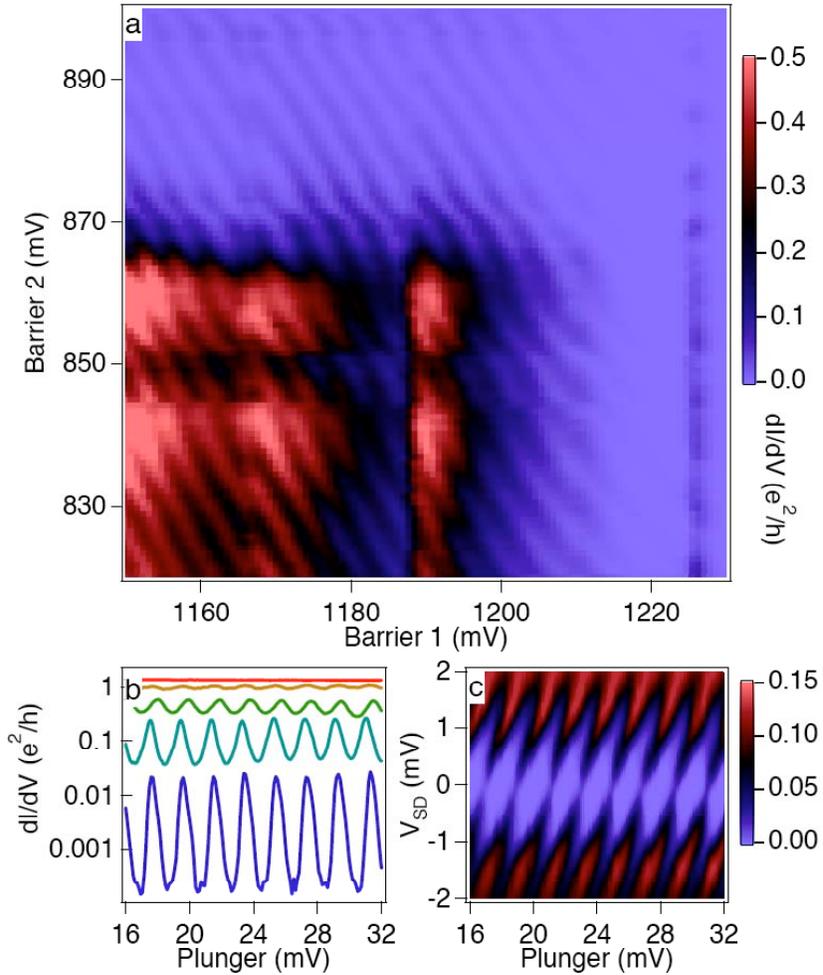

Figure 2a. Differential conductance, *dI/dV*, plotted as a function of barrier gate voltages. Conductance can be suppressed by the application of either gate voltage, leading to the observed corner. Both barrier gates couple capacitively to the carrier densities in the proximal sections of the nanotube and also to the chemical potential of the dot formed between the depletion regions. Thus, near full pinch-off with both gates we observe the emergence of a series of diagonal lines in the 2D plot due to the onset of single electron charging.

2b. *dI/dV*, on a logarithmic scale as a function of the plunger gate for various values of barrier gate voltages. Data measured on a subsequent cool down from those in panels a and c, where the exact position of the corner has shifted slightly in gate voltage. From bottom to top, data are taken at various barrier gate values falling along a diagonal line (with slope approximately 0.8 in the Barrier 1 – Barrier 2 plane) starting from the corner at which conductance is pinched off towards lower gate voltages where conductance is larger. At the highest barrier gate voltages, well-isolated Coulomb blockade peaks are observed; decreasing the barrier gate voltages yields Coulomb oscillations on a high conductance background and eventually open transport without charging effects.

2c. Differential conductance, $dI/dV$, (in units of $e^2/h$) as a function of plunger gate voltage and source-drain voltage, $V_{SD}$, for Barrier 1 (2) = 1200 (880) mV. Coulomb diamonds (regions of suppressed conductance) indicate where the charge on the dot is fixed.

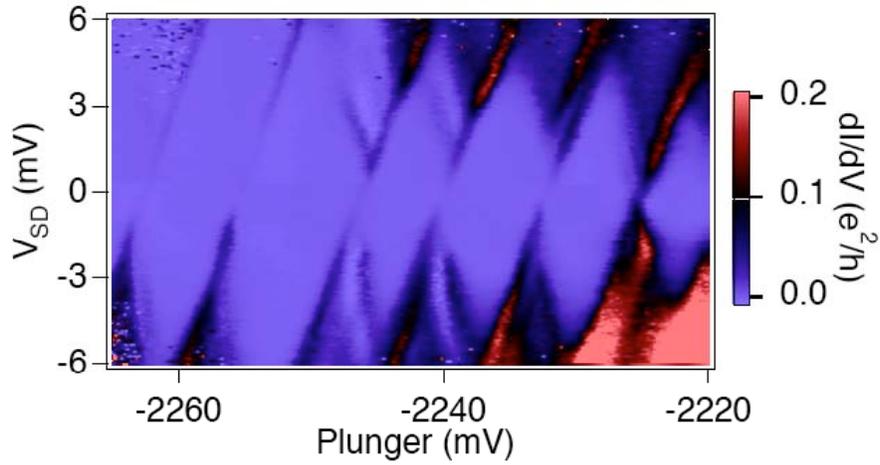

Figure 3. Coulomb diamonds measured on a device using ALD $Al_2O_3$ as the gate dielectric for T ~ 270 mK. Excited states are visible outside of the diamonds as off-resonance conduction. Gates were ~150 nm wide with ~150 nm spacing. Barrier 1(2) = -993 (-2337) mV, Backgate = +16 V. This device is ambipolar, and here operated in the electron-doped regime.

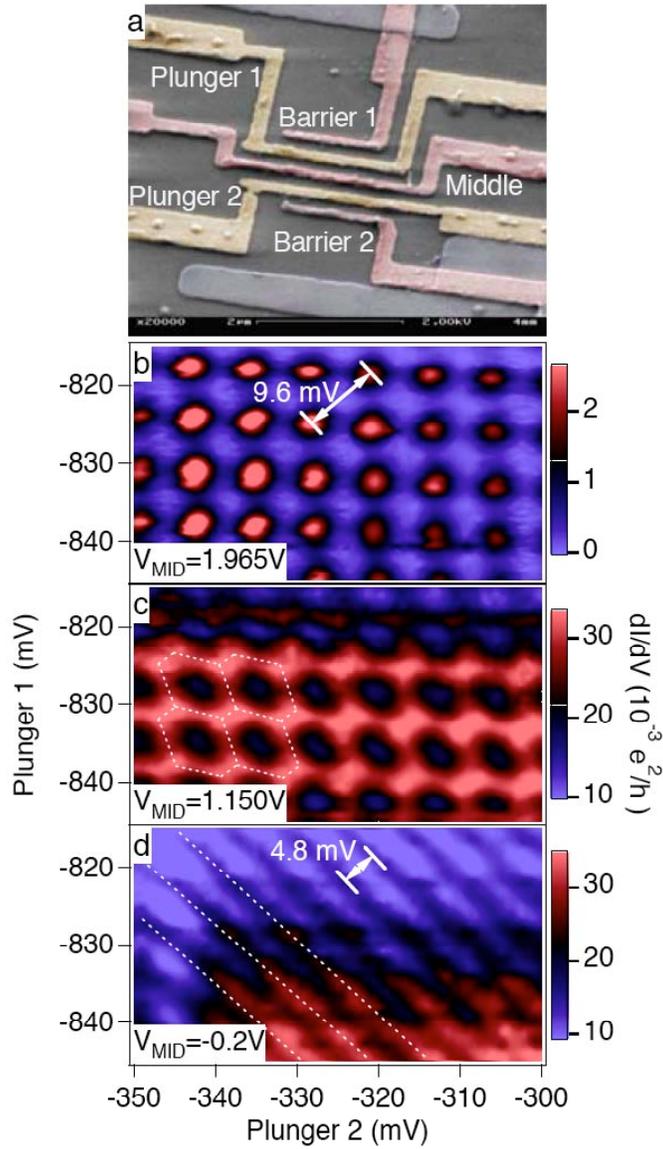

Figure 4a.  Colored SEM image of a five-gate carbon nanotube device lithographically similar to that measured. Pd contacts are visible at the top and bottom, under the $SiO_2$ insulator. The nanotube itself is not visible.  Gates used to form dots are colored red while plunger gates which tune dot energy levels are colored yellow. Scale bar = 2 μm.

4b.  *dI/dV* (colorscale) as a function of two plunger gate voltages. Barrier 1 = 389 mV, Barrier 2 = 1077 mV, Middle Gate voltage is indicated on the figure.  Temperature = 1.5K, ac excitation = 53 μV.  For weak interdot coupling, high-conductance points appear on a regular array corresponding to resonant alignment of energy levels between the two dots with the Fermi levels of the leads. Note the low overall conductance.

4c. At intermediate interdot coupling (for lower middle gate voltage), cross-capacitance and tunneling between dots splits the degeneracy points, giving the familiar hexagonal double-dot charging diagram (dashed lines). Dotted while lines serve as guides to the eye.

4d. For strong coupling of the double dot (lowest middle gate voltage) the two plunger gates together on a single effective dot, producing single-dot charge states separated by diagonal stripes (dashed lines). Note the factor ~2 change in peak period between (b) and (d)